\documentstyle[12pt]{article}

\begin{document}
\renewcommand{\theequation}{\thesection.\arabic{equation}}
\thispagestyle{empty} \vspace*{-1.5cm} \hfill {\small SU-ITP-99-28}\\
\hspace*{\fill} {\small NSF-ITP-99-049}\\[8mm]

\setlength{\topmargin}{-1.5cm} \setlength{\textheight}{22cm}
\begin{center}
{\large  Negative Energy, Superluminosity and Holography} \\

\vspace{2 cm} {\large Joseph Polchinski}\\ Institute for Theoretical
Physics, University of California\\ Santa Barbara, CA 93106-4030 \\
\vspace{.2 cm} {\large Leonard Susskind and  Nicolaos Toumbas}\\ Department
of Physics,
Stanford University\\ Stanford, CA 94305-4060 \\ \vspace{3cm}

\begin{abstract}
The holographic connection between large $N$ Super Yang Mills
theory and gravity in anti deSitter  space requires unfamiliar
behavior of the SYM theory in the limit that the curvature of the
AdS geometry becomes small. The paradoxical behavior includes
superluminal oscillations and negative energy density. These
effects typically occur in  the SYM description of events which
take place far from the boundary of AdS when the signal from the
event arrives at the boundary.  The paradoxes can be resolved by
assuming a very rich collection of hidden degrees of freedom of
the SYM theory which store information but give rise to no local
energy density. These degrees of freedom, called precursors, are
needed to make possible sudden apparently acausal energy momentum
flows. Such behavior  would be impossible in classical field
theory as a consequence of the positivity of the energy density.
However we show that these effects are not only allowed in quantum
field theory but that we can model them in free quantum field
theory.

\end{abstract}
\vspace{1cm} {\it March 1998}
\end{center}

\setcounter{equation}{0}
\section{Introduction and Review}

That there is some sort of  holographic \cite{1,2} correspondence
between maximally supersymmetric $SU(N)$ Yang Mills theory and
supergravity or string theory on $AdS_5 \times S^5$ \cite{3,4,5}
has been established beyond reasonable doubt.  In this paper we
will assume the correspondence in the strongest sense, namely,
that SYM theory and IIB string theory are equivalent for all
values of $N$ and gauge coupling constant.

Equivalence between theories in different dimensions immediately
raises questions about how detailed bulk information in one theory
can be completely coded in lower dimensional degrees of freedom.
Despite the large amount of evidence that we have for the AdS/CFT
correspondence, there is not yet any direct translation of the
configurations of one theory to the other.  We will see that the
existence of such a dictionary requires behavior for the large $N$
limit of SYM theory which seems very unusual and even unphysical,
but we will argue that it is neither.

The parameters of the SYM theory are the gauge coupling constant
$g$ and the number of colors $N$. These are related to the radius
of curvature $R$ of the $AdS_5 \times S^5$, the string length
scale $l_s$ and the string coupling $g_s$ by
\begin{eqnarray}
g_s &=& g^2 \cr R &=& l_s (N g^2)^{1/4}\ .
\end{eqnarray}
The 5 and 10 dimensional Newton constants are given by
\begin{eqnarray}
G_5 &=& G_{10}/R^5 \cr G_{10}&=& g_s^2 l_s^8.
\end{eqnarray}
Throughout we will neglect constants of order one.

The features of the correspondence that are most relevant to our
discussion are the following:

{\bf 1.}  Certain local gauge invariant SYM operators correspond
to bulk supergravity fields evaluated at the boundary of the AdS
space.  That is, the expectation value of the SYM operator is
determined from the boundary value of the field, as explained more
fully in section~3.  For example the energy momentum tensor
$T_{\mu \nu}$ of the SYM theory corresponds to the metric
perturbation  $\gamma_{\mu \nu}$. Similarly $F_{\mu \nu} F^{\mu \nu}$
corresponds to the dilaton field $\phi$.

{\bf 2.} The ultraviolet-infrared connection \cite{6} relates the
short wave length ultraviolet modes of SYM theory to the
supergravity modes near the boundary of AdS space. To be more
precise let us introduce ``cavity coordinates" in which the AdS
metric $ds^2$ is written in terms of a dimensionless metric $dS^2$
and the radius of curvature $R$
\begin{eqnarray}
ds^2 &=& R^2 dS^2 \nonumber\\ dS^2 &=& \biggl({1+r^2 \over
1-r^2}\biggr)^2 dt^2 - \biggl({2 \over 1-r^2}\biggr)^2 (dr^2 + r^2
d \Omega^2)\ .
\end{eqnarray}
The coordinates $t,r$ are dimensionless. The center of AdS means
the point $r=0$. Near a point of the boundary at $r=1$ the metric
has the form

\begin{equation}
ds^2=R^2\left[{1\over z^2}(dt^2 - dz^2 - dx^i dx^i) \right]
\end{equation}
where $z=1-r$ and $x^1,x^2, x^3$ replace the coordinates of the
3-sphere.
For our purposes the metric (1.4) is to be regarded as a local
approximation to the cavity metric.  It is true, but irrelevant to
our purposes, that the same metric also gives an exact description
of a patch of AdS space. In any case we will call these   the
half­plane coordinates.

The SYM theory will be
thought of as living on the dimensionless unit sphere $\Omega$
times the dimensionless time $t$. All quantities such as energy,
distance and time in the SYM theory are regarded as
dimensionless. To relate them to corresponding bulk quantities the
conversion factor is $R$. Thus for example, a time interval
$\delta t$ corresponds to a proper interval $R \delta t$ in the
bulk theory. Similarly an energy $E_{\rm SYM}$   in the SYM theory
is related to the bulk energy by $E_{\rm SYM} = E_{\rm bulk} R$.

The UV--IR connection states that supergravity degrees of freedom
at $1-r = \delta$ correspond to SYM  degrees of freedom with
wavelength $\sim \delta$. The UV--IR connection is at the heart of
the holographic requirement that the number of degrees of freedom
should be of order the area of the boundary measured in Planck
units.  It also suggests that physical systems near the center of
the AdS space should be described by modes of the SYM
theory of the longest wavelength, that is the homogeneous modes on
$\Omega$.

{\bf 3.} The existence of a flat space limit \cite{7,8}. This
limit involves $N \to \infty$ but  is not the 't Hooft
limit in which $g^2 N$ and the energy are
kept fixed. The flat space limit is given by
\begin{eqnarray}
g^2 &\to& {\rm fixed} \cr N&\to& \infty\ .
\end{eqnarray}
In addition, all energy scales are kept fixed in string units.
This means that the dimensionless SYM energy scales like $R$, or
using eq.~(1.1)
\begin{equation}
E_{\rm SYM} \to N^{1/4}\ .
\end{equation}
As argued in  \cite{7,8} an S-matrix can be defined in this limit
in terms of SYM correlation functions (see refs. \cite{9} for related developments). We will outline the
construction but the reader is referred to the references for
details.

A bulk massless particle of  energy $k$ is described by a SYM
excitation of energy $\omega$ given by
\begin{equation}
\omega = Rk = (N g^2)^{1/4}l_s k\ .
\end{equation}
In order to obtain definite kinematics in the flat space limit, the scattering must occur in a known position due to the position-dependence of the metric. 
Therefore, we require the particles to collide within a space-time
region called the ``lab".  The lab is centered at $t=r=0$ and has
a large but fixed size $L$ in string units. At the end we may take $L/l_s$ as big as we like.\footnote {For
generic wavepackets the analysis in refs.~\cite{7,8} shows that
they grow as $N^{1/8}$ due to geometric optics effects, but for
simplicity we imagine special packets chosen to intersect in a
volume of order $N^0$.  The size and duration of any collision
process is determined by the external energies and so is of order
$N^0$.} In terms of dimensionless coordinates the lab dimensions
are
\begin{equation}
\delta t \sim \delta r \sim L/R\ .
\end{equation}
Since $L$ is fixed in string units
\begin{equation}
\delta t \sim \delta r \sim (Ng^2)^{-1/4}\ .
\end{equation}

Creation and annihilation operators for emitting and absorbing
particles at the AdS boundary can be defined. In order that the
particles pass through the lab they must be emitted at time
$t_{\rm in} \approx -{\pi / 2}$. The collision process lasts for a
fixed time in string units which means a dimensionless time of
order $ N^{-1/4}$. Thus in  the dimensionless time of the SYM
theory the duration of the collision becomes negligible as $N \to
\infty$. The outgoing particles will arrive at the boundary
at time $t_{\rm out} \approx {\pi / 2}$. The graviton creation
operators corresponding to this situation are given by
\begin{equation}
A_{\rm in} = \int dt\, d^3 x\, T_{\mu \nu}(x,t) e^{i \omega t}
G(x-x_0,t+ {\pi/ 2 })
\end{equation}
where the integration is over the boundary and $x_0$ is the point
from which the graviton is emitted. Annihilation operators $A_{\rm
out}$ at $t = \pi/2 $ are defined in a similar manner. In order to
make sure that the particles pass through the origin the functions
$G$ in eq.~(1.8) must not be too sharply peaked at $x_0$. We refer
the reader to \cite{7,8} for a discussion of this point.

Let us consider a state involving a packet of gravitons emitted at
$t = -\pi /2$ from the boundary at point $x_0$, in a uniform
state on the $S^5$. The packet is very well concentrated in the
dimensionless coordinates $t,r,\Omega$. In order to translate this
into the SYM theory we need to compute the gravitational field of
such a source.  Fortunately this has been done in \cite{10}. The
result is an AdS generalization of the Aichelburg Sexl metric and
like that metric, it is described by a shock wave that propagates
in the bulk with the particle.  The thickness of the shock in
dimensionless coordinates tends to zero with the spread of the
packet. The intersection of the shock wave with the boundary forms
a 2-sphere or shell which expands away from the point $x_0$ with
the speed of light. Both in front of the shell and behind it
$ \langle T_{\mu\nu} \rangle$ vanishes. The shell expands to its maximum
size at
$t=0$ and then contracts to the antipodal point at $t = +\pi /2$.
Thus the expectation value of the SYM energy momentum tensor
has its support on such a
moving shell and is zero everywhere else.

The description given above is somewhat surprising in view of the
UV--IR connection. We might have expected that in the SYM
description the energy would be transferred from the short wave
length modes of the field theory to long wave lengths as the
graviton moves toward $r=0$. In this event the sharp features of
the shell should have dissipated.  However the energy  stays
concentrated in a thin shell whose thickness tends to zero with
$N$. This in itself is somewhat puzzling.

\setcounter{equation}{0}
\section{History of a Collision}

Paradoxes become apparent when we consider the collision of two
packets. The packets are emitted from points $x_1,x_2$ at time
$t=-\pi/2 $. The first thing to notice is that from the viewpoint
of the boundary SYM theory the behavior of the system after $t=0$
becomes infinitely sensitive to the details of the emission
process. As an example suppose $x_1$ and $x_2$ are separated by an
angle of 90 degrees on the 3-sphere $\Omega$. If the two particles
are emitted at the same time they will reach $r=0 $ simultaneously
and collide. But suppose the emission processes are separated by
time $\epsilon \ll 1$. In this case the arrivals will be separated
by a time \it{in string units} \rm of order $ \epsilon
(Ng^2)^{1/4}$. This means that if $\epsilon > (Ng^2)^{-1/4}$ the
particles will miss each other and pass essentially unscattered.
On the other hand if $\epsilon < (Ng^2)^{-1/4}$ a collision will
take place leading to a very different final state. In other words
shifting the parameters of the emission process by tiny amounts
will lead to large differences in the outcome.

Consider a process in which two packets of fixed energy in string
units are emitted from diametrically opposing points in such a way
that they pass through the lab. In the SYM theory the process
starts out as a pair of expanding thin shells of energy. At time
$t=0$ the shells meet. Now from what was said above, one might
expect the evolution just after  $t=0 $ to be supersensitive to
to the initial conditions. However this is not true. In fact the
two shells just pass through one another without any apparent
interaction. The reason is that as $R \to \infty$, points near the
boundary are so far (again in string units ) from the sources that
the gravitational  field equations linearize.  The energy-momentum
continues to be concentrated on the thin shells which are now
contracting toward the points antipodal from where they
originated. Not only is $ \langle T_{\mu\nu} \rangle$ zero
everywhere off the shells but so are all other SYM fields that
correspond to classical supergravity fields.

We now come to a critical question. The vanishing of $\langle
T_{\mu \nu} \rangle$ in classical field theory would indicate that
the local state of the system is vacuum-like. In other words all
fields or functionals of fields supported off the shells should
have their vacuum values. As we shall see, this can not be true in
the quantum theory. We will find that after the shells pass
through each other the region between them must be excited away
from the local vacuum configuration despite the fact that the
expectation value of the energy density (as well as the value of
every SYM field which corresponds to classical supergravity)
vanishes.

A simple illustrative example is the case where the two packets
are prepared so as to collide head-on at $t=r=0$. Assume the
particles have a fixed energy which is  much larger than the
10-dimensional Planck mass. In this case they will form a
10-dimensional  Schwarzschild black hole. Not all the energy will
go into the black hole but much of it will continue to propagate
as gravitational bremsstrahlung. According to the assumption of a
flat space limit \cite{7,8} the percentage of energy trapped in
the black hole can be calculated in the flat space limit
\cite{11}. The black hole quickly becomes spherically symmetric
and then decays by Hawking evaporation. The entire history of the
black hole lasts a  fixed time in string units and therefore a
dimensionless time which tends to zero like $N^{-1/4}$.

How does the creation and evaporation of the black hole affect the
metric and other supergravity fields at the AdS boundary? The
answer is that it doesn't, at least at first. In fact the
supergravity fields do not respond until light has had a chance to
propagate from $r=0$ to the boundary. The evaporating black hole
sends out a spherically symmetric signal which  arrives at the
boundary at $t= \pi /2$. The arrival of the signal is very sudden,
occupying a time $ \delta t \sim (Ng^2)^{-1/4}$. At this time the
entire boundary suddenly ``lights up" with a spherically symmetric
distribution of energy which in total equals the mass of the black
hole. In other words a fraction of the energy originally stored in
the collapsing shells very quickly flows and is redistributed into
a homogeneous component. We will call this phenomenon ``light-up."

This behavior seems extremely bizarre. The instantaneous
rearrangement of energy appears to violate causality. However this
may not be so. To better understand it we will  describe an
analogous example involving a sudden flow of electric charge.
Consider an example in which initially there is a concentration of
charge in some region $R_1$. At time $t=0$ the charge is found to
disappear from $R_1$ and reappear at $R_2$ which is outside the
forward light cone of $R_1$. To see how this can happen, imagine a
wire connecting $R_1$ and  $R_2$. The wire is full of electrons
and positive ions so that it is electrically neutral. Now we
prearrange observers at each point of the wire so that at $t=0$
they move   each electron slightly toward $R_2$. The result is
a sudden appearance of charge at the ends with no charge density
ever occurring anywhere else.  If there was already a charge at
$R_1$ it would be cancelled by the new charge at that point. The
net result would be a sudden redistribution of charge.

Two ingredients are necessary for such behavior. The first is that
the current vector $j^{\mu}$ be spacelike. Since the charge
density on the wire is always zero it is clear that the current is
purely in the spacelike direction.  This also means that in some
frame the charge density was negative. This of course is not a
difficulty since charge density  can be either positive or
negative.

The other ingredient is {\it prearrangement}. The physical
conditions along the wire must include agents with synchronized
clocks that are  instructed in advance to  act simultaneously.

The sudden flow of energy requires the same two ingredients. In
order that the energy is rearranged so suddenly the flux of energy
$\langle T^{0i} \rangle$ must be much larger than the energy
density itself. This means that the energy density can be made
negative by a Lorentz boost. However, unlike electric charge,
energy is not allowed to be negative in SYM theory. Since in
classical SYM theory the energy density is positive this kind of
flow of energy is absolutely forbidden in the classical theory.
However quantum theory allows local negative energy densities as
long as they are (over)compensated by nearby positive energy
density \cite{12}. In the next section we will analyze this in
more detail and show that the bounds as in ref.~\cite{12} are
consistent with the behavior required of the SYM theory.

A second puzzle concerns locality. Let us suppose that before
light-up the region between the separating  shells is physically
indistinguishable from the vacuum of the SYM theory. By this we
mean that all expectation values of functionals of fields in this
region are identical to their vacuum values. Then the light-up is
impossible. To see this we consider a point $(x,t)$ on the
boundary just after light-up. The point is not near the points
where the shells are localized. The SYM Heisenberg equations of
motion can be used to express the energy density at this point in
terms of local fields at a time just before light-up. Furthermore,
causality requires that the only fields that can be involved are
in the region between the shells where we have assumed vacuum
conditions. It follows that the energy density at $(x,t)$ must be
the same as for the vacuum, that is, zero.

The resolution of these paradoxes, assuming the correspondence is
really as strong as we believe, must be that the region between
the shells at time $0 < t< \pi/2$ must not be vacuum-like even though
the expectation value of the energy-momentum tensor vanishes. Thus
we are forced to postulate that in a region in which $\langle
T_{\mu \nu}\rangle = \langle F^{\mu \nu}F_{\mu \nu}\rangle
=\ldots=0$, the vacuum is excited to a non-vacuum-like local state
which provides the precursor for the later event that we called
light-up. The precursor fields must play the role of the
prearranged agents which simultaneously move charges in the
electric example. Furthermore there must be a very rich manifold
of such precursor configurations. To see this suppose we change
the initial emission parameters by a small amount of order
$N^{-1/4}$. As we have seen this can lead to a very large change
in the results of the collision. For example such a change can
cause an increase of the impact parameter so that a peripheral
grazing collision results. In this case the particles may get
deflected through a small angle. Again, the news of the collision
does not arrive at the boundary until $t= \pi /2$. As before the
energy must suddenly rearrange but this time the result is not a
spherically symmetric component but a new pair of localized small
shells at shifted positions. Also as before, the information must
be  locally stored in a configuration with vanishing expectation
value for the energy momentum tensor. Evidently all the local
physical processes that can take place in the lab are coded in
precursor configurations.

\setcounter{equation}{0}
\section{Gravitational Wave}

{\bf The Principal.}

In this section  we will consider a simplified example
in which a gravitational wave propagates radially outward
from $r \sim 0$. We begin with the wave in the linearized theory in
which the field equations are treated to lowest order in the
deviations from AdS space. Assume that the wave is in one of the
lowest spherical harmonics on the 3­-sphere. In half­plane
coordinates the plane fronted wave has the form
\begin{equation}
\gamma_{\mu \nu}(z,x,t) = \xi_{\mu\nu}R^2z^2f(z,t)
\end{equation}
where $\gamma_{\mu \nu}(z,x,t)$ is defined by
\begin{equation}
ds^2=R^2\left[{1\over z^2}(dt^2 - dz^2 - dx^i dx^i) \right]+ \gamma_{\mu \nu}(z,x,t)dx^{\mu}dx^{\nu}
\end{equation}
and $\xi_{\mu\nu}$ is a transverse traceless polarization tensor
with nonvanishing components in the $x$ directions. The
polarization tensor is assumed normalized to unity.

According to the AdS/CFT correspondence, the wave makes a
contribution to the SYM energy momentum tensor given by
\cite{13}\cite{14}\cite{15}
\begin{equation}
<T_{ij}> \sim -{R \over G_5}z^{-2} \gamma_{ij}|_{z=0} = - \xi_{ij}{R^3 \over G_5} f(0,t).
\end{equation}
We assume that at some initial time $t_0$ in the past, the
function $f(z,t)$ describes a wave propagating toward $z=0$ and
that it vanishes for $z < t_0$ . Thus at the initial time the
contribution to $<T>$ from the wave vanishes. Furthermore
causality of the bulk theory insures that $f(0, t)$ will remain
exactly zero until $t = 0$. At that time $f(0, t)$ and the SYM
stress tensor begin to oscillate. After a time the wave will be
reflected and the value at $z = 0$ will return to zero
exponentially. Thus far we have considered the gravitational wave
in the linearized approximation. By analogy with terminology
introduced in \cite{12}
the contribution to $<T>$ in this approximation is called the
{\it principal}.

\bigskip
\noindent{ \bf The Interest.}

The nonlinear corrections to the gravitational field equations
give rise to a correction to the metric which in turn corrects the
SYM energy density. Again by analogy with \cite{12} this term is called the
{\it interest}. It is smaller
than the principal by a factor $G_5$. To compute the interest let
us return to cavity coordinates. Consider any spherical\footnote{Near the boundary, the gravitational wave looks like a plane sheet with uniform energy density. The correction to the metric at a given point should be the same with that of a large spherical distribution of the same energy density.}
distribution of energy which is non vanishing only for $r< r_0$.
Then, as in flat space, the gravitational field at $r> r_0$ is
completely determined to be that of a neutral non­rotating black
hole with the same total energy. The metric of a black hole in AdS
is given in Schwarzschild-­like coordinates by
\begin{equation}
ds^2=R^2\left[(1+b^2-{2MG_5 \over R^2 b^2})dt^2 - (1+b^2-{2MG_5
\over R^2 b^2})^{-1}db^2 - b^2 d\Omega^2\right]
\end{equation}
where $b$ is the radial coordinate. The coordinates $b$ and $r$
are related by
\begin{equation}
1+b^2= {(1+r^2)^2\over(1-r^2)^2}.
\end{equation}
Near the boundary this becomes
\begin{equation}
b={1 \over {1-r}}.
\end{equation}
Thus near the boundary the time--time component of the metric has
the form
\begin{equation}
g_{00} = g_{00}^{AdS} - 2MG_5(1-r)^2.
\end{equation}
This gives a contribution to $\gamma_{00}$ given by
\begin{equation}
\gamma_{00} = -2MG_5(1-r)^2 = -2MG_5 z^2.
\end{equation}
Using the time--time component of eq(3.3) we find the interest
\begin{equation}
<T_{00}^{int}> \sim MR.
\end{equation}

   Thus to compute the interest we need to
compute the energy M stored in the gravitational wave. Our goal
will be to compute the energy in terms of data on the boundary.
This will facilitate comparison with the boundary CFT. For our
purposes, it is convenient to set
\begin{equation}
\gamma_{\mu \nu} = {h_{\mu \nu} \over z^2} = \xi_{\mu \nu} R^2
{\Phi(z, t) \over z^2}.
\end{equation}
We work in the gauge $h_{z \mu} = 0$. In the case when $\xi_{\mu
\nu}$ is traceless with no timelike components, the constraints
from the $z\mu$ components of the Einstein equations are
automatically satisfied. The linearized equations for $h_{ij}$
reduce to the following differential equation for $\Phi(z, t)$
\begin{equation}
\partial_{z}^2\Phi - {3 \over z} \partial_z\Phi - \partial_t^2\Phi = 0.
\end{equation}
The corresponding $1+1$ dimensional Lagrangian density is obtained
to be
\begin{equation}
{ \cal L} = {R^2 \over 2G_5z^3} \left [ {(\partial_t\Phi)}^2 -
{(\partial_z\Phi)}^2 \right ].
\end{equation}
This is just the Lagrangian density of a minimally coupled
massless scalar field in AdS. Using the equation of motion, we can
write the energy stored in the gravitational wave as
\begin{equation}
{M} = {R^2 \over  2G_5}\int{ {dz \over z^3}
\left [{(\partial_t\Phi)}^2 - \Phi \partial_t^2 \Phi \right ]}.
\end{equation}

We now look for solutions that fall like $z^4$ near the boundary
so that $\gamma_{\mu \nu}$ falls like $z^2$. If we set
\begin{equation}
\Phi = z^2  { \chi }(z)  e^{-i\omega t},
\end{equation}
then ${ \chi}$ satisfies a standard Bessel differential equation
\begin{equation}
z^2 \partial_z^2 { \chi} + z \partial_z { \chi} - (4 -
z^2\omega^2) { \chi} = 0
\end{equation}
with solution $J_2(\omega z)$. Thus, the most general solution can
be written as follows
\begin{equation}
\Phi (z, t) = {z^2 \over 2} \int{d\omega \phi(\omega) J_2(\omega z)
e^{-i\omega t}} + cc.
\end{equation}
Near the boundary, ${J_2(\omega z)} \sim {{(\omega z)}^2}$, and so
\begin{equation}
f(\omega) = \omega^2 \phi(\omega)
\end{equation}
is the Fourier transform of the boundary data.

It is a simple exercise to compute the energy, and, therefore, the
interest in terms of the boundary data. Using (3.13), (3.16)
and the  orthogonality relations for Bessel functions
gives
\begin{equation}
\left < T^{int}_{00} \right > \sim {R^3 \over  G_5} \int{d\omega
{\left|f(\omega)\right|^2 \over \omega^3}}.
\end{equation}

There are two things to note about the interest. First is that it
is completely featureless having neither space nor time
dependence. The second is that in a certain sense it can be made
arbitrarily small. To see this let us compare the interest  with
the principal obtained in eq(3.3)
\begin{equation}
<T_{ij}> \sim - \xi_{ij} {R^3 \over G_5} f(0,t) = -\xi_{ij} {R^3 \over 2G_5}
\int{d\omega f(\omega) e^{-i\omega t}} + cc.
\end{equation}
Defining
\begin{eqnarray}
P&=&<T_{ij}> \cr I&=& T_{00}^{int}
\end{eqnarray}
we note that the ratio $I/P^2$ is given by
\begin{equation}
{I \over P^2} \sim   G_5/R^3.
\end{equation}
Now using eqs(1.1) and (1.2) we can rewrite this in terms of SYM
quantities
\begin{equation}
{I \over P^2} \sim   1/N^2.
\end{equation}
The point is that if we hold fixed the principal then the interest
goes to zero like $1/N^2$. Therefore in the large $N$ limit the
interest becomes negligible.

To summarize, the AdS/CFT correspondence requires the following
behavior for the SYM energy momentum tensor. For $t<0$ the energy
density and pressure are constant with respect to spatial and
temporal position. For large $N$ they are vanishingly small $\sim
N^{-2}$. At $t=0$ the stress tensor begins to oscillate
simultaneously over all space with magnitude of order unity. After
a time the wave is reflected and the oscillations cease.

\section{Squeezed States in Field Theory}
The superluminal behavior and non--positivity of the energy-momentum
tensor are incompatible with classical field theory. However as we will see
in this section, they are not only compatible with quantum field theory but
can even
be found in the theory of free fields.   For
notational simplicity we will use the theory of $N^2$ free scalar
fields $\phi_{nm}$.  Thus, we will be able to obtain a field
theoretic model for the gravitational wave we have just described.

Conformal invariance requires that we use the improved form of the
energy-momentum tensor, which in four dimensions is given by
\begin{equation}
T_{\mu \nu} = Tr({2 \over 3}
{\partial_{\mu}{\phi}}{\partial_{\nu}{\phi}} - {1 \over
6}\eta_{\mu \nu} {(\partial_{\sigma}{\phi})}^2 - {1 \over 3} \phi
{\partial_{\mu} {\partial_{\nu} {\phi}}} + {1 \over 12} \eta_{\mu
\nu} \phi \partial^2 {\phi}).
\end{equation}
In this form, the energy-momentum tensor is traceless and
satisfies the usual conservation law. In what follows we
normal-order the energy-momentum tensor so that all creation
operators appear to the left of all annihilation operators. This
is equivalent to setting the vacuum energy to zero.

Consider the squeezed state \footnote{The terminology ``squeezed
state'' originated in the quantum optics literature. The squeezing
refers to the shape of the oscillator phase space probability
distributions.}
\begin{equation}
\left| \psi \right> = \exp \left [{{1 \over 2} \int{d^3 \vec{k}
d^3 \vec{k'}\ F(\vec{k}, \vec{k'})\
a^\dagger_{mn}(\vec{k})a^\dagger_{nm}(\vec{k'})}} \right] \left| 0
\right>
\end{equation}
for the particular choice
\begin{equation}
F(\vec{k}, \vec{k'}) = F(\vec{k}) \delta^3(\vec{k} + \vec{k'})
\end{equation}
such that $F(\vec{k}) = F(-\vec{k})$. We consider the case when
$F$ is small; then,
\begin{equation}
\left< \psi \right| T_{\mu \nu} \left| \psi \right> = T^{(1)}_{\mu
\nu} + T^{(2)}_{\mu \nu}
\end{equation}
with the first piece being linear in $F$ and the second piece
being quadratic in $F$.

The effect linear in $F$ is obtained from contracting two
annihilation operators in $T_{\mu \nu}$ with two creation
operators in $\left | \psi \right >$, or vice versa with $\left <
\psi \right|$. The gauge indices in each pair are contracted among
themselves, and so $T^{(1)}$ is of order $N^2$. All timelike
components, $T^{(1)}_{0 \mu}$, are zero as one can easily verify.
The non zero diagonal components are given by
\begin{equation}
T^{(1)}_{ii} = { N^2 \over 2} \int{{d^3\vec{p} \over
\omega_{\vec{p}}} (p_i^2 -{1 \over 3}{\omega_{\vec{p}}}^2)
F(\vec{p}) e^{-2i\omega_{\vec{p}}t}} + cc,
\end{equation}
and the off-diagonal components
\begin{equation}
T^{(1)}_{ij} = { N^2 \over 2} \int{{d^3\vec{p} \over
\omega_{\vec{p}}} p_i p_j F(\vec{p}) e^{-2i\omega_{\vec{p}}t}} + cc.
\end{equation}
As expected $T^{(1)}$ is homogeneous, traceless and oscillatory in
time. In all, five independent components are non-zero as in the
case of the gravitational wave studied above. Therefore, we
identify this piece with the principal.

The only non-vanishing piece which is quadratic in $F$ has the
form $\left <F \right| T \left | F\right>$.  It receives
contributions from terms in $T_{\mu \nu}$ with one creation and
one annihilation operator only. It is also of order $N^2$ and
homogeneous. In particular,
\begin{equation}
   T^{(2)}_{00} = { N^2(2\pi)^3} \int{{d^3\vec{p}  \omega_{\vec{p}}}
\left|F(\vec{p})\right|^2}
\end{equation}
is positive and time-independent. We identify this piece with the
interest.

We are now ready to compare with the results found in the previous
section. First, we note that the ratio
\begin{equation}
{T_{2} \over {{(T_{1}})}^2} \sim {1 \over \ N^2}
\end{equation}
has the same scaling with $N$ as before. Moreover, we can obtain a
consistent relation between the Fourier transform of the boundary
data, $f(\omega)$, and the angle average of $F(\vec{p})$. If we
compare $T^{(1)}$ with the principal, we find
\begin{equation}
f \sim { F  \omega^3  }.
\end{equation}
Now, if we substitute $f/ \omega^3$ for $F$ in eq(4.29), we
obtain the same connection between interest and
principal as given in eq(3.18).

We proceed now to study the time dependence of the principal. We
wish to show that it can be oscillatory for a certain period of
time and then vanishingly small. Take
\begin{equation}
F(\vec{p}) = \alpha {p_3^2 \over p^3} {\left (p^2 \lambda \right
)}^{n/2} e^{-\lambda p^2}
\end{equation}
as an example. Here, $\alpha$ has units of mass and $\lambda$ has
units of inverse mass squared. We consider the case for which the
number $n$ is even. We focus on a particular non-zero component.
For example,
\begin{equation}
T^{(1)}_{33} = { 4 \alpha {\lambda}^{-{3 \over 2}} \pi N^2 \over
45} \int^{+\infty}_{-\infty}{dx x^{(2 + n)} e^{-x^2} e^{2ix{t \over \sqrt{\lambda}}}}.
\end{equation}
This in turn is equal to
\begin{eqnarray}
T^{(1)}_{33} &=& { 8 \alpha {\lambda}^{-{3 \over 2}} \pi N^2 \over
45} {{\left(\sqrt{\lambda} \over 2i\right)}^{( 2 + n)}}
\partial^{(2 + n)}_t {\int^{+\infty}_{-\infty}{dx e^{-x^2} e^{2ixt\over
\sqrt{\lambda}}}}\cr
&=& { 8 \alpha {\lambda}^{-{3 \over 2}} \pi N^2
\over 45}{{\left(\sqrt{\lambda} \over 2i \right)}^{( 2 + n)}}
\partial^{(2 + n)}_t \left(\sqrt{\pi}e^{-{t^2 \over \lambda}}\right).
\end{eqnarray}
The final result is a polynomial in $t$ times a Gaussian. This
means that the principal is oscillatory for some period of time near $t=0$
(depending on $\lambda$) and, then, it vanishes
exponentially as $t \rightarrow \infty$.

It is interesting to consider gravitational waves which propagate
along a direction which is not perpendicular to the boundary.
Suppose the wave vector has components in the $(x^1,z)$ plane. In
this case the wave fronts  do not simultaneously arrive at the
boundary. In fact the boundary data itself becomes a wave
propagating in the $x^1$ direction. Furthermore the wave has both
group and phase velocity greater than $1$. It is superluminal! To
see how this happens in the CFT we can compute the principal with
the delta function in eq(4.25) replaced by $\delta (\vec{k} + \vec{k'}
-\vec{q})$
where $\vec{q}$ lies along the $x^1$ direction. The resulting principal
is easily computed and forms a superluminal wave. Actually it is
not necessary to do any further calculation to see the bulk--boundary agreement in this case. The gravitational wave
propagating in the $z,x^1$ direction can be obtained by an
AdS-Lorentz transformation from the original wave propagating
along $z$. This maps into a conformal transformation of the
boundary. Since the field theory we are using is conformally
invariant the agreement for the transformed wave follows from the
agreement in the original case.

In addition to being superluminal, the wave in the $x^1$ direction
also has nonvanishing oscillating time--time component so that the
energy density  oscillates from positive to negative \cite{12}.

\setcounter{equation}{0}
\section{Precursors}
The profile of the gravitational wave obviously carries
information. We would like to understand what precursor
degrees of freedom carry that information, particularly during the
time $t<0$ before the wave arrives at the boundary. The interest
$T_{\mu\nu}^{int}$ is completely featureless  in both space and
time and cannot be relevant here. Furthermore all bulk fields
vanish within a neighborhood of the boundary. This means that the
local SYM fields that can be identified with boundary values of
bulk fields also vanish.

Evidently   the precursors are nonlocal. In
the free field example a convenient example is given by
\begin{equation}
\Phi (x,x') =({2 \over 3}\partial_{\mu} \partial'_{\nu}
-{1 \over 6}\eta_{\mu \nu}\partial_{\sigma}\partial'^{\sigma} - {1
\over 3} \partial'_{\mu}\partial'_{\nu})
<\phi(\vec{x},t)_{mn}\phi(\vec{x}',t')_{nm}>|_{t=t'}
\end{equation}
where the expectation value is taken in the squeezed state. This
quantity obeys a wave equation with respect to $t$ and $(\vec{x} - \vec{x}')/2$.
Furthermore at $\vec{x} - \vec{x}' = 0$ it is given by $<T_{\mu \nu}>$. At early
times when $<T_{\mu \nu}>$ vanishes $\Phi (x,x')$ is nonzero for
$|\vec{x} - \vec{x}'| \approx 2|t|$. In other words the precursor becomes
increasingly nonlocal the further the wave is from the boundary.
This is of course a manifestation of the IR/UV connection
\cite{6}. Furthermore at any time  $\Phi (x,x')$ has the same
information as the function $F$ that characterizes the squeezed
state.

\setcounter{equation}{0}
\section{Conclusion}

The main purpose of the free field model 
is to demonstrate that some of the odd superluminal and negative
behavior of $T$ predicted by the AdS/CFT correspondence is
consistent with the principles of quantum field theory. It is
somewhat surprising that the free field model works so well and
captures the detailed relation between principal and interest. We
don't really know why this is so but it seems to be part of a
pattern. For example, free field theory describes the
thermodynamics of AdS black holes correctly apart from a well
known factor of $3/4$ in the effective number of degrees of
freedom. More exactly, free field theory agrees with AdS/CFT
predictions for the two and three point functions of chiral
primaries. We suspect that if the model scalar field theory is
replaced by   free SYM theory the numerical relation between
interest and principal may be exact as a consequence of the
non-renormalization theorems of the two and three point functions.

Perhaps the most interesting result of this paper is
the identification of nonlocal precursor fields such as
\begin{equation}
\phi_{mn}(x) \phi_{nm}(x').
\end{equation}
These fields would have to be modified in the interacting SYM
since they are not gauge invariant as they stand. A candidate
would be
\begin{equation}
\phi_{mn}(x) \phi_{rs}(x')W_{mr}W'_{sn}
\end{equation}
where $W$ and $W'$ are Wilson lines along two paths connecting the
points $x,x'$. In fact the entire object (6.2) can be thought of
as a single Wilson loop. This suggests that the nonlocal
precursors which code local bulk information are expectation
values of Wilson loops of size dictated by the UV/IR connection.

\section{Acknowledgements}

We would like to thank  V. Balasubramanian, G. Horowitz, N.
Itzhaki, A. Peet, M. Peskin,  S. Ross and S. Shenker for helpful
conversations. This work
was supported in part by NSF grants PHY98-70115, PHY94-07194 and PHY97-22022.

%
%
%

%
%
%
\setcounter{equation}{0}
%
%
\end{document}